\documentstyle[twoside,fleqn,espcrc2,epsfig]{article}

\def\slashchar#1{\setbox0=\hbox{$#1$}           
   \dimen0=\wd0                                 
   \setbox1=\hbox{/} \dimen1=\wd1               
   \ifdim\dimen0>\dimen1                        
      \rlap{\hbox to \dimen0{\hfil/\hfil}}      
      #1                                        
   \else                                        
      \rlap{\hbox to \dimen1{\hfil$#1$\hfil}}   
      /                                         
   \fi}                                         %

\newcommand{\ifig}[1]{\mbox{\epsfig{file=#1,height=40mm,width=75mm}}}

\def\bc{\begin{center}}
\def\ec{\end{center}}
\def\be{\begin{equation}}
\def\ee{\end{equation}}
\newcommand{\chilim}{\lim_{m \rightarrow 0}}
\newcommand{\ba}{\begin{eqnarray}}  
\newcommand{\ea}{\end{eqnarray}}

\newcommand{\nn}{\nonumber}

\newcommand{\msbar}{\overline{\mbox{\scriptsize MS}}}

\newcommand{\bea}{\begin{eqnarray}}
\newcommand{\eea}{\end{eqnarray}}

\newcommand{\as}{\alpha_s} 

\newcommand{\W}{\hat{   W}}
\newcommand{\U}{\hat{ U}}
\newcommand{\gammaz}{\hat{\gamma}^{(0)T}}

\newcommand{\AmS}{{\protect\the\textfont2
  A\kern-.1667em\lower.5ex\hbox{M}\kern-.125emS}}

\title{Matrix Elements without Quark Masses on the Lattice}

\author{A.~Donini\address{Dep. de Fisica Teorica, Univ. Autonoma de Madrid, 
          Cantoblanco, E-28049 Madrid, Spain.},
        V.~Gim\'enez\address{Dep. de Fisica Teorica and IFIC, 
          Univ. de Valencia, 
          E-46100 Burjassot, Valencia, Spain.},
        L.~Giusti\address{Department of Physics, Boston University
         Boston, MA 02215 USA.}\thanks{Talk presented by L.~Giusti.
         This research was 
         supported in part under DOE grant DE-FG02-91ER40676.},
         G.~Martinelli.\address{Dip. di Fisica, Univ. di Roma ``La Sapienza''
          and INFN-Sezione di Roma, I-00185 Roma, Italy.}}
       
\begin{document}

\begin{abstract}
\vspace{-.3cm}
We introduce  a new parameterization of four-fermion  matrix elements which
does not involve quark masses and thus allows a reduction of   
systematic uncertainties in physical amplitudes. As a result the 
apparent quadratic dependence of $\epsilon'/\epsilon$ on $m_s(\mu)$ is 
removed. To simplify the matching
between  lattice and continuum renormalization schemes, we
express our results in terms of   Renormalization Group Invariant
$B$-parameters which  are renormalization-scheme and scale 
independent. As an application of our proposal, matrix elements 
of  $\Delta I=3/2$ 
and SUSY $\Delta F =2$ ($F=S,C,B$) four-fermion operators  have been computed.
\vspace{-0.5cm}
\end{abstract}

\maketitle
\section{Introduction}
\label{sec:intro}
\vspace{-.3cm}
Since  the original proposals of using lattice QCD to study hadronic 
weak decays~\cite{CMP,BGGM,bernardargo}, 
substantial theoretical and numerical progress has been made:
the main theoretical aspects of the renormalization of composite four-fermion
operators  are well understood \cite{boc,shape1,4ferm_teo}; the calculation 
of  $K^0$--$\bar K^0$ mixing has reached a level of accuracy which 
is unpaired by any other approach \cite{Laur};  
increasing precision has been gained in the determination of the
electro-weak penguin amplitudes necessary to  the prediction
of the CP-violation parameter $\epsilon^\prime/\epsilon$ 
\cite{gupta_bp,NoiDELTAS=2,ds2susy}; finally matrix  elements of $\Delta S=2$ 
operators which are relevant to study FCNC effects in SUSY models have been 
computed~\cite{NoiDELTAS=2,ds2susy}. Methods, symbols and results 
reported in this talk are fully described in \cite{NoiDELTAS=2,ds2susy}.
\vspace{-.6cm}
\section{Matrix elements without quark masses}
\label{sec:definitions}
\vspace{-0.3cm}
The analysis of $K^0-\bar K ^0$ mixing with the most
general $\Delta S =2$ effective Hamiltonian requires   
the knowledge of matrix elements $\langle\bar K^0|O_i |K^0 \rangle$ of  
parity conserving parts of the following operators 
\bea 
O_1 &=& \bar s^\alpha \gamma_\mu (1- \gamma_{5} ) d^\alpha \ 
\bar s^\beta \gamma_\mu (1- \gamma_{5} )  d^\beta ,  \nn \\ 
O_2 &=& \bar s^\alpha (1- \gamma_{5} ) d^\alpha \ 
 \bar s^\beta  (1- \gamma_{5} )  d^\beta ,  \nn \\ 
O_3&=& \bar s^\alpha  (1- \gamma_{5} )  d^\beta  \ 
 \bar s^\beta   (1- \gamma_{5} ) d^\alpha ,  \label{eq:ods2} \\ 
O_4 &=& \bar s^\alpha  (1- \gamma_{5} ) d^\alpha \  
\bar s^\beta  (1 + \gamma_{5} )  d^\beta ,  \nn \\ 
O_5&=& \bar s^\alpha  (1- \gamma_{5} )  d^\beta \ 
 \bar s^\beta (1 +  \gamma_{5} ) d^\alpha . \nonumber
 \eea
On the lattice, matrix elements of weak four-fermion operators 
are computed from first principles. But, following the common lore, 
they are usually 
given in terms  of the  so-called $B$-parameters which measure 
the deviation  of their values from those obtained in the Vacuum 
Saturation Approximation (VSA). 
For operators in (\ref{eq:ods2}), the
$B$-parameters are usually defined as
\bea\label{eq:bparC} 
\langle  \bar K^{0} \vert   O_{1} (\mu) \vert K^{0} 
\rangle &=& \frac{8}{3} M_{K}^{2} f_{K}^{2} B_{1}(\mu) \, ,  \\
\langle  \bar K^{0} \vert  O_{i} (\mu) \vert K^{0} \rangle  &=&
\frac{C_i}{3} \left( \frac{ M^2_{K} f_K }{ m_{s}(\mu) + 
m_d(\mu) }\right)^{2} B_{i}(\mu)\, ,\nonumber
\eea
where $C_i=-5,1,6,2$ for ($i=2,\dots , 5$).
In (\ref{eq:bparC}), $\langle \bar K^0|O_1 |K^0 \rangle$ is
parameterized in terms of well-known experimental  quantities and $B_1(\mu)$ 
($B_K(\mu)\equiv B_1(\mu)$). On the contrary, $\langle \bar K^0|O_i |K^0 \rangle$ 
($i=2,\dots ,5$) depend quadratically on the quark masses in (\ref{eq:bparC}),
while they are expected to remain finite in the chiral limit and depend only  
linearly on the quark masses. Contrary to $f_{K}$, $M_{K}$,  etc., 
quark masses  can not be  
directly measured by experiments and the present accuracy  
in their determination is still rather  poor. 
\begin{table}[tbhp]
\vspace{-0.5cm}
\setlength{\tabcolsep}{.10pc}
\newlength{\digitwidth} \settowidth{\digitwidth}{\rm 0}
\catcode`?=\active \def?{\kern\digitwidth}
\caption{\it{Matrix elements in GeV$^4$ at the renormalization scale 
$\mu = 2$~GeV in the RI scheme obtained with the new 
parameterization and the conventional one 
in ref.~\cite{NoiDELTAS=2} on the same set of data.}}
\begin{tabular}{||c|c|cc||c|cc||}\hline\hline
      & \multicolumn{2}{c}{New}& &\multicolumn{2}{c}{Old}& \\
\hline
$\langle  O_i \rangle$ &$\beta=6.0$&$\beta=6.2$& 
& $\beta=6.0$  &$\beta=6.2$& \\
 &this work& this work & &\cite{NoiDELTAS=2}
&\cite{NoiDELTAS=2}& \\
\hline \hline
$\langle  O_1 \rangle$& 0.012(2) & 0.011(3) & & 0.012(2) & 0.011(3) & \\ 
$B_1$   &                   0.70(15) & 0.68(21) & & 0.70(15) & 0.68(21) & \\
\hline
$\langle  O_2 \rangle$&-0.079(10)&-0.074(8) & &-0.073(15)&-0.073(15)& \\ 
$B_2$                     & 0.72(9)  &0.67(7)   & & 0.66(3)  & 0.66(4)  & \\
\hline
$\langle  O_3 \rangle$&0.027(2)  & 0.021(3) & &0.025(5)  &0.022(5)  & \\ 
$B_3$                     &1.21(10)  & 0.95(15) & & 1.12(7)  & 0.98(12) & \\
\hline
$\langle  O_4 \rangle$& 0.151(7) & 0.133(12)& &0.139(28) &0.133(28) & \\ 
$B_4$                     & 1.15(5)  & 1.00(9)  & & 1.05(3)  & 1.01(6)  & \\
\hline
$\langle  O_5 \rangle$& 0.039(3) & 0.029(5) & &0.035(7)  &0.029(7)  & \\ 
$B_5$                     & 0.88(6)  & 0.66(11) & & 0.79(6)  & 0.67(10) & \\
\hline
$\langle  O^{3/2}_7 
\rangle$                  & 0.019(2) & 0.011(3) & & 0.020(5) & 0.014(5) & \\ 
$B^{3/2}_7$               & 0.65(5)  & 0.38(11) & & 0.68(7)  & 0.46(13) & \\
\hline
$\langle  O^{3/2}_8 
\rangle$                  & 0.082(4) & 0.068(8) & & 0.092(19)& 0.087(19)& \\ 
$B^{3/2}_8$               & 0.92(5) &0.77(9) & & 1.04(4) & 0.98(8) & \\
\hline \hline
\end{tabular}
\label{tab:summary}
\end{table}
 \begin{table}[htb]
\vspace{-0.8cm}
\centering
\caption{\it{RGI Matrix elements in GeV$^4$ computed as in Eq.~(\ref{eq:BRGI}) 
with $\alpha_s^{n_f=4}$.}}
\label{tab:summaryrgi}
\begin{tabular}{||c|c|c||}\hline\hline
$\langle  O^{RGI}_i\rangle$ &$\beta=6.0$&$\beta=6.2$\\
\hline \hline
$\langle  O^{RGI}_1 \rangle$ & 0.017(3) & 0.016(4) \\ 
\hline
$\langle  O^{RGI}_2 \rangle$ & -0.051(7) & -0.048(6) \\ 
\hline
$\langle  O^{RGI}_3 \rangle$ & 0.005(7) & -0.004(7) \\ 
\hline
$\langle  O^{RGI}_4 \rangle$ & 0.072(3) & 0.063(6) \\ 
\hline
$\langle  O^{RGI}_5 \rangle$ & 0.043(3) & 0.032(5) \\ 
\hline \hline
\end{tabular}
\vspace{-0.8cm}
\end{table}
Therefore,
whereas for $O_1$ we introduce $B_{K}$ 
as an alias of the matrix element,  by using (\ref{eq:bparC}) we 
replace each of the "SUSY" matrix elements with 2 unknown quantities, i.e. 
the $B$-parameter and $m_{s} + m_d$.
To overcome these problems, 
we  propose the following new parameterization of $\Delta S =2$ operators 
\bea\label{eq:FURBAdef}
\langle  \bar K^{0} \vert   O_{1} (\mu) \vert K^{0} 
\rangle &=& \frac{8}{3} M_{K}^{2} f_{K}^{2} B_1(\mu) ,  \\
\langle  \bar K^{0} \vert   O_{i} (\mu) \vert K^{0} \rangle  &=&
M_{K^*}^{2} f_{K}^{2} \tilde{B}_i(\mu).\nn
 \eea
The $\tilde B_{i}(\mu)$-parameters are still 
dimensionless quantities and can be computed on the lattice
by studying appropriate ratios of three- and 
two-point functions \cite{ds2susy}. By simply using them, we have eliminated  
any fictitious reference to the quark masses, hence reducing the 
systematic errors on the corresponding 
physical amplitudes. An alternative parameterization, not
used here, which can be useful in the future is reported 
in \cite{ds2susy}.
The VSA and $B$-parameters are also used for matrix elements of  
operators which enter the $\Delta S =1$ effective Hamiltonian. 
Notice that this "conventional" parameterization is the only 
responsible for the apparent 
quadratic dependence of $\epsilon'/\epsilon$ on the quark masses.
This introduces a 
redundant source of systematic error which can be 
avoided by parameterizing the matrix elements in terms of measured
experimental quantities and 
therefore {\it a better determination of the strange quark mass 
$m_s(\mu)$ will not improve our theoretical knowledge of 
$\epsilon'/\epsilon$}. 
In this work we have computed the matrix
elements $\langle\pi|O^{3/2}_{i}|K\rangle$ of the four fermion operators
$O^{3/2}_{i}$ ($i=7,8,9$) which contribute to the $\Delta I=3/2$ sector of 
$\epsilon'/\epsilon$. In the chiral limit 
$\langle\pi\pi|O^{3/2}_{i}|K\rangle$  
can be obtained,  using  soft pion theorems,  from  
$\langle\pi^{+}|O^{3/2}_{i}|K^{+}\rangle$. For degenerate quark 
masses, $m_{s}=m_{d}=m$, and in the chiral limit,  we find
\ba\label{eq:softpion}
\chilim \langle\pi^{+}|O^{3/2}_{7}|K^{+}\rangle & = & 
 - M_{\rho}^{2} f_{\pi}^{2} \chilim \tilde{B}_5(\mu)\nonumber\\
\chilim \langle\pi^{+}|O^{3/2}_{8}|K^{+}\rangle & = &
 - M_{\rho}^{2} f_{\pi}^{2} \chilim \tilde{B}_4(\mu)\nonumber\\
\chilim \langle\pi^{+}|O^{3/2}_{9}|K^{+}\rangle & = &
\frac{8}{3} M_{\pi}^{2} f_{\pi}^{2} \chilim B_1(\mu)\; .  \nonumber
\ea
\vspace{-0.7cm}
\section{Renormalization Group Invariant Operators}
\label{sec:RGI}
\vspace{-0.3cm}
Physical amplitudes can be written as 
\be 
\langle F \vert {\cal H}_{eff}\vert I \rangle = 
\langle F \vert \vec{O}(\mu) \vert I \rangle \cdot \vec{C}(\mu) 
\, , \label{wope} 
\ee
where $\vec{ O}(\mu) \equiv
( O_1(\mu),\dots,  O_N(\mu))$ is the operator basis 
(for example the one in (\ref{eq:ods2}) for the 
$\Delta S = 2$) and
$\vec{C}(\mu)$   the corresponding Wilson coefficients
represented as a column vector.
$\vec C(\mu)$ is  expressed in terms of its counter-part, 
computed at a large scale $M$, 
through the renormalization-group evolution matrix  $\W[\mu,M]$
\be 
\vec C(\mu) = \W[\mu,M] \vec C(M)\, , \label{evo} 
\ee
where the initial conditions $\vec C(M)$, are obtained by 
perturbative matching 
of the full theory to the effective one 
at the scale $M$ where all the heavy particles have been removed. 
$\W[\mu,M]$ can be 
written as (see for example \cite{scimemi})
\bea
\W[\mu,M]  = \hat M[\mu] \U[\mu, M] \hat M^{-1}[M] \, , 
 \label{monster} \eea
where $\U = (\alpha_s(M)/\alpha_s(\mu))^{(\gamma^{(0)T}_O/2\beta_0)}$ is the 
leading-order evolution matrix
and $M(\mu)$ is a NLO matrix defined in \cite{scimemi}.
The Wilson coefficients $\vec{C}(\mu)$ and the renormalized operators 
$\vec{ O}(\mu)$ 
are usually defined in a given scheme, at a fixed renormalization scale $\mu$, 
and they depend on the renormalization scheme and scale in such a way 
that only $H_{eff}$ is scheme and scale independent. 
To simplify the matching procedure, we propose a Renormalization
Group Invariant (RGI) definition of Wilson coefficients and 
composite operators 
which generalizes what is usually done for $B_K$ 
and for quark masses. We define 
\be\label{eq:CRGI}
\hat{w}^{-1}[\mu] \equiv  \hat M[\mu] \left[\as (\mu) \right]^{
      - \gammaz_O / 2\beta_{ 0}}\, , 
\ee
and using  Eqs.~(\ref{monster}) and (\ref{eq:CRGI}) we obtain
\be
\W[\mu,M]  = \hat{w}^{-1}[\mu]\hat{w}[M]\; .  
\ee
The effective Hamiltonian (\ref{wope}) can be written as 
${\cal H}_{eff}=\vec O^{RGI} \cdot \vec C^{RGI}$, where
\be\label{eq:monster2}
\vec C^{RGI}      =  \hat{w}[M] \vec C(M)\; , \;
\vec O^{RGI}     =   \vec{ O}(\mu) \cdot \hat{w}^{-1}[\mu]\; .
\ee
{\it $\vec C^{RGI}$ and  $\vec O^{RGI}$ are scheme and scale independent at  
the order we are working. Therefore
the effective Hamiltonian is splitted in terms which are individually 
scheme and scale independent}. This procedure is 
generalizable to any effective weak Hamiltonian.
The RGI $\tilde{B}$-parameters can be defined as 
\be\label{eq:BRGI}
{\tilde B}_i^{RGI} = \sum_{j} \tilde{B}_j(\mu)   w(\mu)_{ji}^{-1}\, .  
\ee
\vspace{-.7cm}   
\section{Numerical results}
\label{sec:numeri}
\vspace{-0.3cm}
All details concerning the extraction of matrix elements from 
correlation functions and the computation of 
the non-perturbative renormalization constants of  
lattice operators can be found in \cite{4ferm_teo,NoiDELTAS=2,ds2susy}. 
In this talk we report the results obtained in \cite{ds2susy}. 
The simulations have  been performed at $\beta = 6.0$ 
(460 configurations) and   $6.2$  (200 configurations) 
with the tree-level Clover
action, for several values of the quark masses and for different 
meson momenta. The main results we have 
obtained for $\Delta S =2$ and $\Delta I=3/2$ matrix elements and their
comparison with the results in \cite{NoiDELTAS=2} are 
reported in Tables~\ref{tab:summary} and \ref{tab:summaryrgi}.
It is interesting to note, as expected from chiral perturbation theory, 
that matrix elements of $\Delta S =2$ SUSY 
operators are enhanced respect to the SM one by a factor $2-12$ at 
$\mu=2$~GeV.
\begin{figure}[tbh]
\vspace{-0.5cm}
\ifig{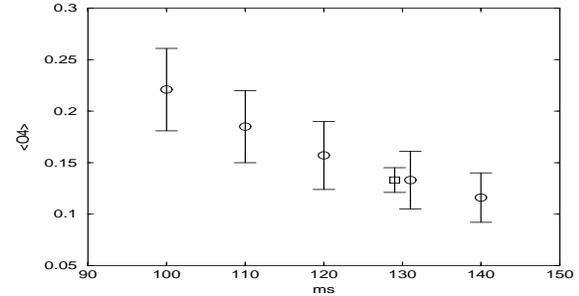}
\vspace{-1.5cm}
\caption{\small{ $\langle  \bar K^{0} \vert O_{4} \vert K^{0} \rangle$ 
(GeV$^4$) 
with the new (square) and old (circle) parameterization as a 
function of the strange quark mass (in $\msbar$ at $\mu=2$ GeV) used 
to obtain the full matrix element from the $B$-parameter.}}
\vspace{-0.7cm}
\label{fig:fig.1}
\end{figure} 
In Figure~\ref{fig:fig.1} we show the strong dependence of 
$\langle  \bar K^{0} \vert O_{4} \vert K^{0} \rangle$ on the 
strange quark mass when the ``conventional'' parameterization 
(\ref{eq:bparC}) is used, to be compared with the result obtained with
the new parameterization. 
The results for the analogous $\Delta C=2$ and $\Delta B =2$ matrix elements 
presented at 
the conference are reported in \cite{NOIDELTAB=2}.      
Although  we have data at two different values of the lattice spacing, the 
statistical errors, and the uncertainties in the extraction of the 
matrix elements,  are too large to enable any extrapolation 
to the continuum limit. 
For this reason, the best estimate of the central values 
of the $B$-parameters can be obtained  
by  averaging the results obtained at the two values of 
$\beta$ \cite{ds2susy}. As far as the errors are concerned we take 
the largest of the two statistical errors. 

\end{document}